\documentclass[10pt,journal,twocolumn]{IEEEtran}

\usepackage{amsmath,amsfonts,amssymb}
\usepackage{array}
\usepackage{textcomp}
\usepackage[hyphens]{url}
\usepackage{graphicx}
\usepackage{cite}
\usepackage{booktabs}
\usepackage{xcolor}
\usepackage[hidelinks]{hyperref}

\makeatletter
\renewcommand\section{\@startsection {section}{1}{\z@}%
  {3.0ex \@plus 1ex \@minus .2ex}%
  {1.0ex \@plus .2ex}%
  {\normalfont\normalsize\scshape\centering}}%
\makeatother

\begin{document}

\title{AI-RAN on NPUs: Baseband Processing Without Baseband Chips}

\author{Shilong~Zhang,~Luping~Xiang,~Jienan~Chen,~and~Kun~Yang%
\thanks{S.~Zhang, L.~Xiang, and K.~Yang are with the State Key
Laboratory of Novel Software Technology, Nanjing University,
Nanjing 210008, China, and with the School of Intelligent Software
and Engineering, Nanjing University (Suzhou Campus), Suzhou
215163, China, e-mail: luping.xiang@nju.edu.cn;
kunyang@nju.edu.cn. \textit{(Corresponding author: Luping~Xiang.)}}%
\thanks{J.~Chen is with the National Key Laboratory of Science
and Technology on Communications, University of Electronic Science
and Technology of China, Chengdu 611731, China (e-mail:
Jesson.Chen@outlook.com).}}

\maketitle

\begin{abstract}
AI-RAN aims to unify artificial intelligence and radio access
network workloads on a shared compute substrate. While this
paradigm has so far been demonstrated primarily on Graphics
Processing Units (GPUs), it remains unclear whether Neural
Processing Units (NPUs), which are AI accelerators optimized for
inference, can also support wireless baseband processing. Here, we provide the
first affirmative answer by resolving the fundamental mismatch
between baseband workloads and NPU architecture. A computational
isomorphism exists: matrix and vector engines NPUs dedicate to
inference inherently cover physical-layer operations. Yet NPU
architectures are natively shaped for dense-tensor AI inference,
not baseband. This architectural mismatch surfaces as opposing
optimization objectives: traditional baseband minimizes arithmetic
operations, whereas NPU performance demands maximizing engine
utilization. We close this gap by reconstructing communication
algorithms onto AI compute primitives, prioritizing engine
utilization over arithmetic count. We validate this with a
complete OFDM transceiver on an Ascend~310B1 edge NPU,
demonstrating end-to-end over-the-air transmission via USRP~X300
at 3.0~GHz.
\end{abstract}

\section*{Introduction}

The integration of artificial intelligence (AI) with radio access
networks, commonly referred to as AI-RAN, is becoming a central
theme in the evolution of 5G and the design of 6G
systems~\cite{pennanen2024_6g,lin2019_5gnr_myths}. AI-RAN is
typically framed around three complementary directions: applying
AI to improve RAN operation and performance (AI-for-RAN),
co-hosting AI and RAN workloads on shared infrastructure
(AI-and-RAN), and exposing AI services through the RAN at the
network edge
(AI-on-RAN)~\cite{kundu2025airan,rathakrishnan2025_airan,open_airan_2025}.
Among these, AI-and-RAN has received growing attention as a
practical route towards infrastructure convergence. For example,
NVIDIA Aerial has demonstrated the concurrent execution of
5G~RAN processing and large language model (LLM) inference on
GH200 servers, where graphics processing unit (GPU) resources are
dynamically partitioned between the two workloads using
Multi-Instance GPU
technology~\cite{kundu2025airan,shah2025_airan_orch,bonati2024_x5g}.
In parallel, Open~RAN platforms have undergone accelerated
commercial deployment in recent
years~\cite{kundu2024hwaccel,polese2023oran,bonati2022_oran,hasabelnaby2024_cran_oran,kirana2026_ml_oran}.
These demonstrations, however, remain rooted in GPU-based
computing and are supported by the mature Compute Unified Device
Architecture (CUDA) software
ecosystem~\cite{nickolls2008cuda}.

\begin{figure*}[!t]
\centering
\includegraphics[width=\textwidth]{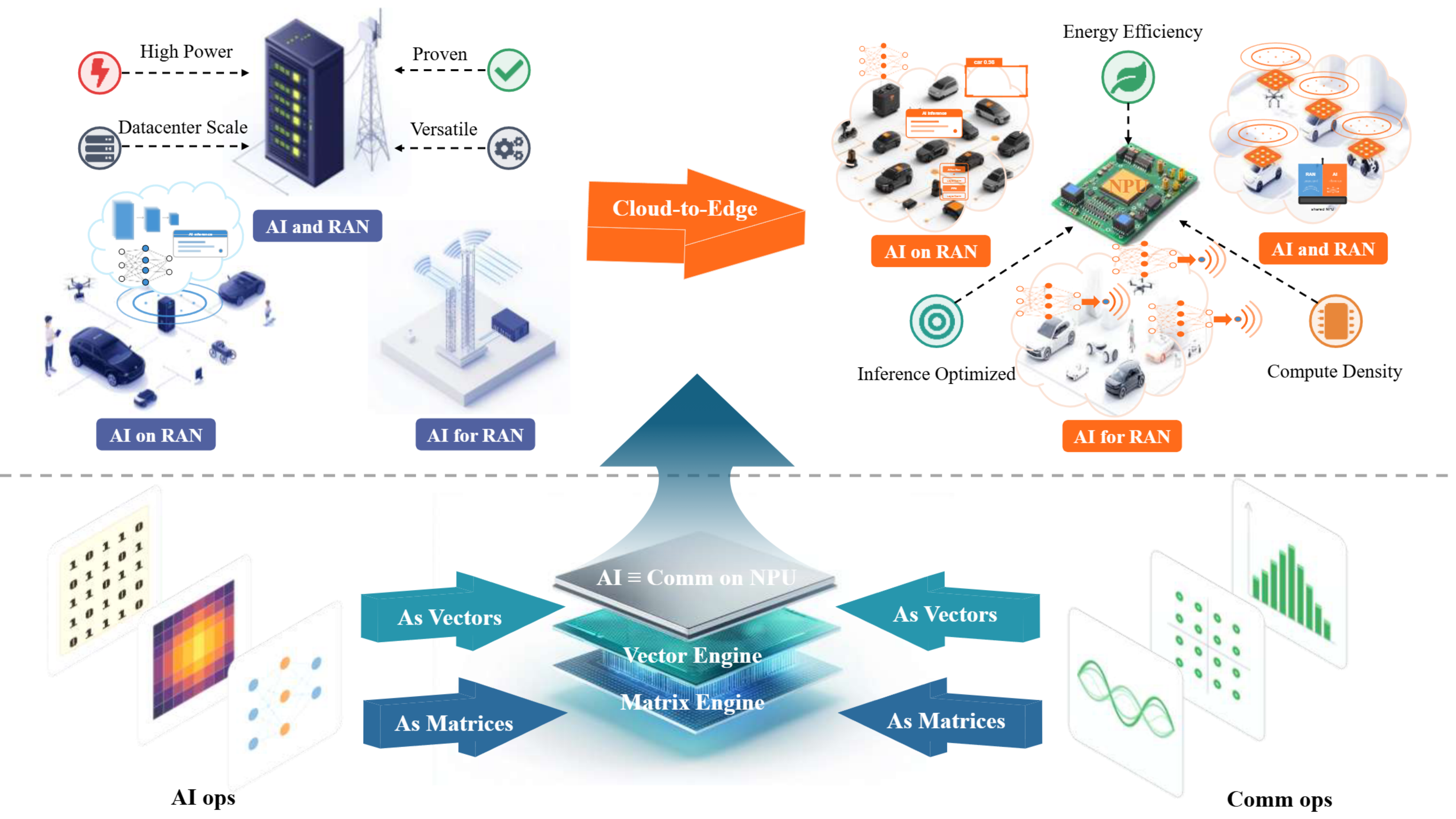}
\caption{\textbf{Two paradigms for realizing AI-RAN.}
\textbf{Top:} GPU-based AI-RAN (left) integrates AI-on-RAN,
AI-and-RAN, and AI-for-RAN within a centralized, high-power server
infrastructure, benefiting from the maturity of GPU software
ecosystems, but limiting energy efficiency and large-scale edge
deployment. NPU-based AI-RAN (right) distributes the same three
functionalities across edge devices by exploiting the energy
efficiency, inference-optimized compute engines, and high compute
density of NPU hardware.
\textbf{Bottom:} enabling principle. AI operators, such as
attention maps, neural networks, and quantized weights, and
communication operators, such as signal waveforms, QAM
constellations, and OFDM spectra, can both be expressed as matrix
and vector primitives executed on common NPU engines. We refer to
this shared computational structure as computational
isomorphism.}
\label{fig:overview}
\end{figure*}

The central premise of AI-RAN is that AI and RAN workloads can
share a common compute substrate. This premise need not be
restricted, in principle, to GPUs. Neural processing units
(NPUs), which are inference-oriented AI accelerators with
architectures distinct from GPUs, provide an alternative
hardware substrate. They are programmed through their own
kernel-level interfaces, such as Huawei AscendC. In this work,
we refer to this class of interfaces as nCUDA, the NPU analogue
of CUDA. Whether such NPU platforms can also support baseband
processing, and thereby realize AI-RAN beyond GPU-centric
servers, has not yet been established. A positive answer would
extend AI-RAN from centralized GPU servers towards distributed
edge NPUs (Fig.~\ref{fig:overview}, top). The key enabling
principle is that AI inference and baseband signal processing
both decompose into matrix and vector primitives executed on the
same accelerator engines (Fig.~\ref{fig:overview}, bottom).

Beyond GPUs, conventional baseband processing has relied mainly
on Field-Programmable Gate Arrays (FPGAs) and Digital Signal
Processors (DSPs). FPGAs provide high throughput and favourable
energy efficiency, but their reliance on Hardware Description
Language (HDL) development slows algorithmic iteration and
deployment~\cite{bishop2019fpga_5gnr}. DSPs, by contrast, offer
C-level programmability, yet lack the dense matrix throughput
required by increasingly AI-enhanced baseband
workloads~\cite{anjum2011dsp_sdr}. NPUs occupy a different point
in this design space. They sacrifice part of the generality of
GPUs in favour of inference specialization, allocating most of
their silicon resources to two dominant computational
primitives: matrix engines for matrix multiplication and vector
engines for element-wise
operations~\cite{npu_efficiency_mdpi,letaief2021_edgeai,chen2020dnn_accel_survey}.
This matrix--vector organization is shared by many contemporary
AI accelerators, including Huawei Da~Vinci Cube/Vector
engines~\cite{davinci_arch}, Qualcomm Hexagon HMX/HVX
units~\cite{codrescu2014hexagon}, Google TPU systolic
arrays~\cite{jouppi2017tpu}, Apple's Neural
Engine~\cite{kasperek2022_apple_m1}, and wafer-scale AI
systems~\cite{lie2019cerebras}. Each of these platforms exposes
its own nCUDA-class kernel-level programming interface, reflecting
a broader architectural convergence documented in recent hardware
surveys~\cite{reuther2022aiml_accel,iwabuchi2025_dl_hpc_accel,wei2025_npu_lpu_survey,markidis2018tensor_cores}.

Here we show that commercial edge NPUs can indeed support
baseband processing. The underlying reason is a
\emph{computational isomorphism} between wireless baseband
operations and AI inference. Core baseband functions such as
transforms, filtering, and channel coding are linear operations
that can be expressed as matrix multiplication, whereas
modulation, equalization, and phase compensation are naturally
represented as element-wise vector operations. Thus, the
fundamental hardware primitives required for baseband processing
are already present in NPU architectures. Hardware coverage alone, however, does not guarantee efficient
execution. Classical baseband algorithms were largely designed to
minimize arithmetic complexity on general-purpose processors. NPU
performance, in contrast, is determined by how effectively the
workload is mapped onto high-throughput matrix and vector engines
rather than low-throughput scalar units. Bridging this gap requires 
re-implementing baseband operators to maximize matrix and 
vector engine utilization.

This mapping exposes two principal challenges. The first is an
\emph{operator-shape mismatch}. NPU engines are optimized for the
large, dense tensors characteristic of AI inference, whereas
baseband operators often span much smaller and more irregular
computational scales. Examples range from small matrices, such as
a $256\times256$ discrete Fourier transform (DFT), to short
vectors produced by filtering and channel estimation, and even to
scalar operations such as the single $\mathrm{atan2}$ evaluation
used in carrier frequency offset estimation. Additional
irregularity arises from butterfly computation graphs, bit-level
XOR sequences and lookup tables, all of which can lead to poor
matrix and vector engine utilization. The second challenge is a
\emph{data-movement bottleneck}. Implementing each baseband
operator as an independent kernel causes intermediate tensors to
move repeatedly between on-chip buffers and off-chip Double Data
Rate (DDR) memory, making memory-transfer latency comparable to,
or even greater than, computation time. Fig.~\ref{fig:methodology}
summarizes the two NPU-native mapping strategies developed in the
Methods section to address these limitations.

Prior work on AI accelerators for the RAN has pursued three
distinct directions, none of which coincides with ours. The
first is GPU-based AI-RAN, realized by NVIDIA
Aerial~\cite{kundu2025airan} and Open~RAN
deployments~\cite{kundu2024hwaccel}, with academic
implementations of GPU-accelerated baseband operators such as
LDPC decoding~\cite{tarver2021_gpu_ldpc}. The second is
custom-designed domain-specific processors for AI-native physical
(PHY) layer~\cite{bertuletti2022mempool_5g}, of which
TensorPool~\cite{tensorpool2026} is a recent representative.
The third is neural-receiver algorithms that replace individual
PHY blocks, treating the physical layer as a deep learning
system~\cite{oshea2017_phy_dl}, such as the model-driven 5G
receiver of~\cite{abdollahpour2025} and NVIDIA's Neural
Receiver~\cite{cammerer2023nrx}; this line of work is supported
by open-source simulation frameworks such as
Sionna~\cite{hoydis2022sionna}. This work differs from all
three: it targets commercial-off-the-shelf (COTS) edge NPUs,
implements the complete Orthogonal Frequency-Division
Multiplexing (OFDM) transceiver chain rather than selected
blocks, and demonstrates the result through over-the-air (OTA)
operation rather than simulation or synthesis.

\begin{figure*}[!t]
\centering
\includegraphics[width=0.7\textwidth]{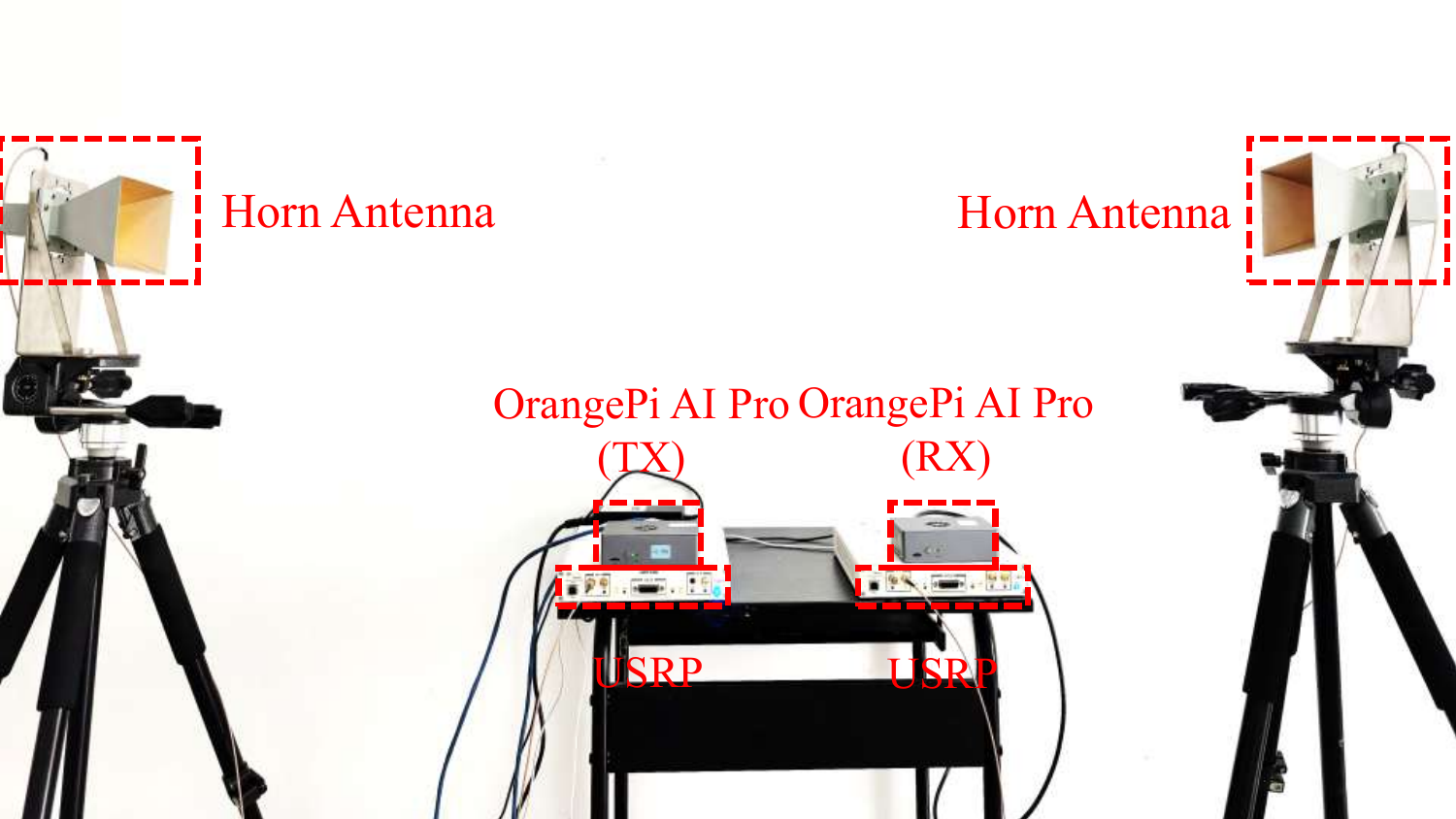}
\caption{\textbf{Experimental setup.} Two Orange Pi AI Pro
single-board computers, each with an on-board Ascend~310B1 NPU,
paired with USRP~X300 software-defined radios and horn antennas
to form a SISO over-the-air link at 3.0~GHz. All baseband
processing executes on the NPUs; the USRPs serve only as RF
front-ends.}
\label{fig:system_setup}
\end{figure*}

\section*{Results}

\subsection*{Over-the-air transmission performance}

% [Finding-Method-Significance]
% Finding: 8 W edge NPU sustains complete OFDM OTA
% Method: two-board USRP X300 setup at 3.0 GHz
% Significance: feasibility — AI inference primitives 能驱动完整物理层
A single 8\,W edge NPU, with no dedicated baseband silicon,
sustains a complete OFDM transceiver over a real wireless link.
To establish this, we built a two-board verification platform
(Fig.~\ref{fig:system_setup}): two Orange Pi AI Pro single-board
computers, each carrying an Ascend~310B1 NPU, paired with
USRP~X300 software-defined radios at 3.0~GHz. The USRP serves
purely as an RF front-end with no baseband processing; all
baseband operations execute entirely on the NPU. Across 200
consecutive frames captured at the receiver, the NPU-resident
chain decoded the transmitted bit stream with a mean post-LDPC
bit error rate of $7.4\times10^{-4}$, with a substantial
fraction of frames decoded without any bit errors (124/200).
Frame synchronization succeeded for 199 of the 200 frames.
Detailed system configuration, OFDM parameters, and operator
implementations are provided in Methods. This result establishes
the feasibility of using a COTS edge NPU as the sole baseband
processor under realistic over-the-air channel conditions,
demonstrating that the architectural primitives built for AI
inference are sufficient to drive a complete physical-layer
transceiver.

\subsection*{Cross-platform latency and energy efficiency}

% [Finding-Method-Significance]
% Finding: 8 W TDP, 4.1 frames/s/W, higher than general-purpose CPUs
% Method: identical chain across four platforms
% Significance: low power + high compute density per watt 对应 edge
%               两个约束(power budget + physical envelope)
At 8\,W Thermal Design Power (TDP), the NPU sustains the OFDM
chain at 4.1\,frames/s/W on transmit, higher than general-purpose
processors at the same workload. To quantify this, we
benchmarked the identical transceiver chain (same algorithms,
same matrix dimensions, same input data) across four platforms
spanning three processor classes: the Ascend~310B1 NPU at 8\,W,
an NVIDIA Jetson Orin~NX edge GPU running cuBLAS at 25\,W (MAXN
mode, 16\,GB), an Intel Core~Ultra~7 270K+ desktop CPU at
125\,W, and the on-chip ARM CPU cores within the same
Ascend~310B1 SoC, following the cross-platform comparison
methodology established in prior edge AI benchmarking
work~\cite{jayanth2024edge_ai_bench}. All baselines execute DFT matrix multiplication
rather than butterfly FFT, ensuring algorithmic equivalence.

End-to-end NPU processing takes 30.7\,ms for the TX chain and
46.2\,ms for the RX chain. The Jetson Orin~NX GPU delivers
65.1\,ms TX and 110.2\,ms RX; the desktop CPU achieves 20.6\,ms
TX latency at 125\,W with a form factor incompatible with
deployment outside a server room. On the same SoC and within
the same 8\,W envelope, the host ARM CPU cores require
255.9\,ms for TX and 536.1\,ms for RX, confirming that the NPU
contribution is the active accelerator rather than the SoC as a
whole. Normalized to power, the NPU delivers 4.1\,frames/s/W
for TX, $6.8\times$ the edge GPU and $10\times$ the desktop CPU
(Fig.~\ref{fig:benchmark}b). Each platform is well-suited to its
intended setting: the desktop CPU for development and offline
processing, the edge GPU for deployments where versatility and
ecosystem maturity dominate. The NPU's combination of low power
draw and high compute density per watt aligns with the two
binding constraints of edge AI-RAN deployment, the power budget
of a battery- or PoE-fed edge node and the physical envelope of
a passively cooled chassis, making the NPU a viable third
operating point for compute-intensive edge workloads.

\subsection*{Per-operator analysis}

\begin{figure*}[!t]
\centering
\includegraphics[width=0.95\textwidth]{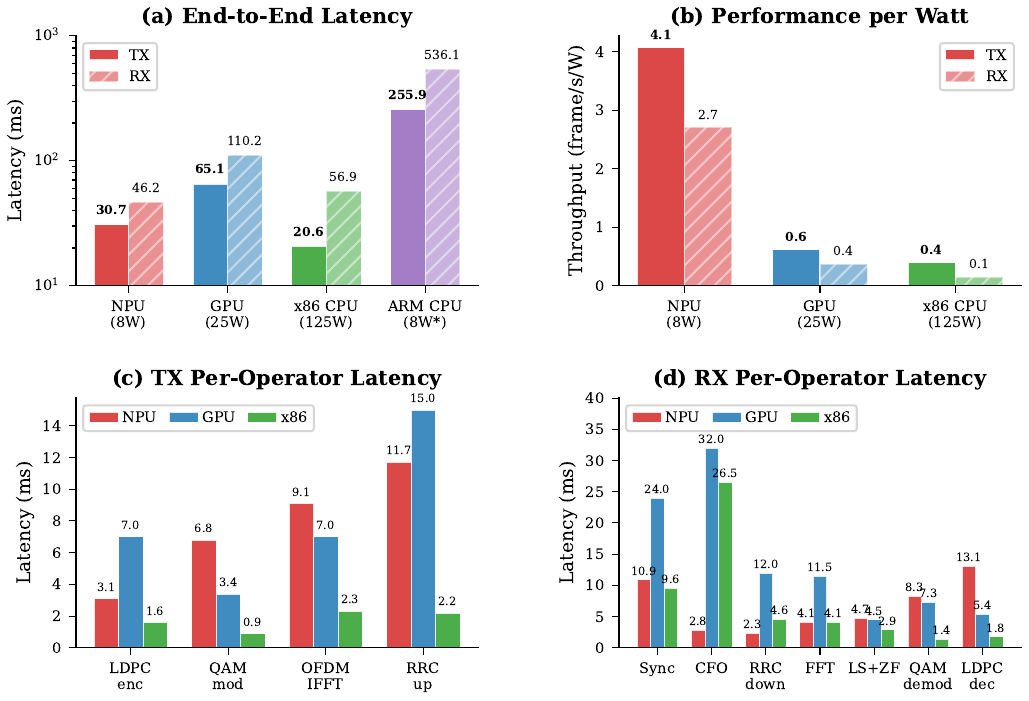}
\caption{\textbf{Cross-platform baseband processing benchmark.}
All platforms execute the identical algorithm with matched matrix
dimensions; the NPU implementation uses C++ host code (pybind11),
while GPU and CPU baselines use Python/CuPy and Python/NumPy
respectively. (\textbf{a})~End-to-end latency on log scale;
\textup{*}shared 8\,W SoC power envelope. (\textbf{b})~Performance
per watt (higher is better); same-board ARM CPU omitted as its
power cannot be isolated from the NPU. (\textbf{c},\textbf{d})~Per-operator
kernel-level latency for the TX and RX chains.}
\label{fig:benchmark}
\end{figure*}

% [Finding-Method-Significance]
% Finding: no single platform dominates; NPU dual-engine balanced
%          coverage
% Method: per-operator latency breakdown
% Significance: balanced coverage > peak single-operator for edge
No single platform dominates across all operators; the
per-operator breakdown reveals distinct strengths across
processor classes. The desktop CPU excels at scalar-intensive
operations such as QAM modulation (0.9\,ms) and CFO compensation,
owing to its high single-thread performance and heavily
optimized BLAS libraries. The GPU achieves the lowest LDPC
decoding latency (5.4\,ms vs.\ 13.1\,ms on the NPU). The NPU,
however, achieves the lowest latency among edge platforms on
Cube-engine operators: RRC filtering (2.3\,ms vs.\ 12.0\,ms on
the GPU), OFDM IFFT (9.1\,ms), and LDPC encoding (3.1\,ms), and
delivers competitive performance across all operator classes
(Fig.~\ref{fig:benchmark}c,d). Baseband processing is a heterogeneous mixture of matrix, vector, and scalar operations, and edge deployment imposes a single power budget on the entire chain rather than on any individual operator. Under this constraint, end-to-end latency at fixed power, rather than peak single-operator latency, determines the achievable frame rate. The 6.8$\times$ end-to-end performance per watt advantage in Fig.~\ref{fig:benchmark}b reflects this: the GPU's lower LDPC-decoder latency does not translate into an end-to-end advantage when amortized over the rest of the chain at the platform's higher power draw, and the CPU's strong single-thread arithmetic does not translate into a deployable form factor at 125\,W. The NPU's dual-engine architecture, by covering both Cube-engine and Vector-engine operators within an 8\,W envelope, matches the workload structure of edge AI-RAN.

\section*{Discussion}

The performance-per-watt advantage observed in our benchmarks
reflects an alignment between edge AI-RAN workloads and NPU
hardware characteristics. Edge AI-RAN deployments are dominated
by forward-pass inference: base stations and edge devices
typically run trained models locally, while the baseband chain
itself is also a forward-pass workload of dense matrix and vector
arithmetic. Edge-class NPUs such as the 310B1 used here dedicate
most of their transistor budget to matrix-vector throughput,
which is well-matched to this regime. GPUs remain the established
choice for AI-RAN where computational versatility, large memory
capacity, and ecosystem maturity dominate the deployment
decision, particularly in datacenter-scale settings. The two
platform classes are complementary: NPUs offer a favorable
operating point when the deployment constraint is throughput
within a fixed thermal envelope.

\begin{figure*}[!t]
\centering
\includegraphics[width=\textwidth]{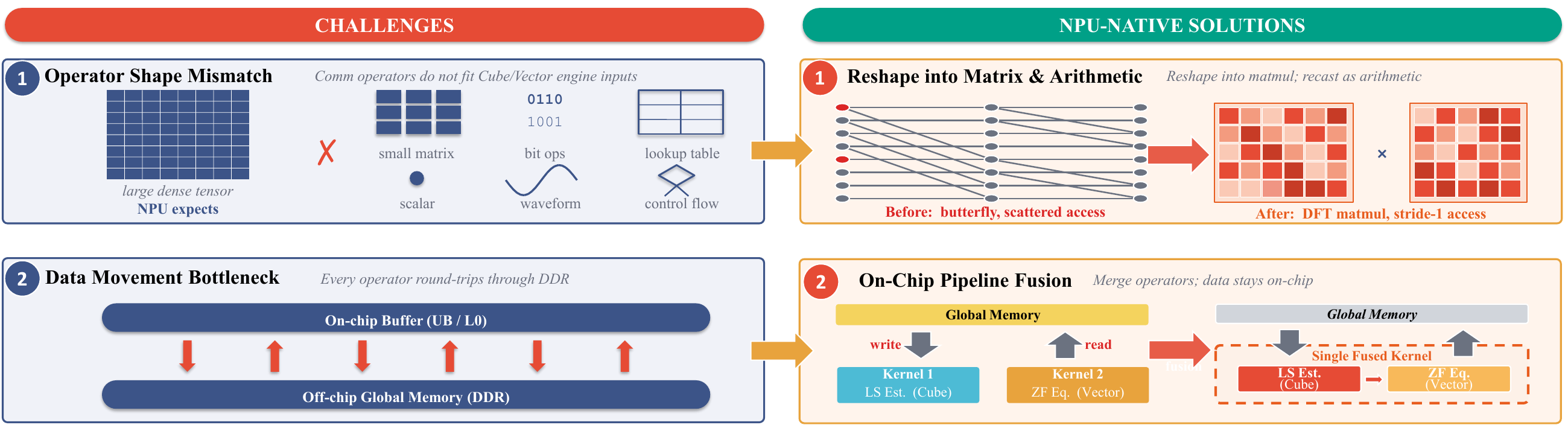}
\caption{\textbf{Two challenges of mapping baseband to NPU and
the two NPU-native solutions that address them.}
\textbf{Challenge~1 (Operator Shape Mismatch):} comm operators do
not fit the dense tensor inputs of Cube and Vector engines, they
appear as small matrices, bit operations, lookup tables, scalars,
waveforms, or control flow. \emph{Solution~1: Reshape into Matrix
and Arithmetic.} Restore matrix multiplication form for linear
operations (e.g., FFT butterfly $\rightarrow$ DFT matmul, with
bit-reversed access transformed into stride-1 access) and recast
nonlinear operations as pure batch arithmetic.
\textbf{Challenge~2 (Data Movement Bottleneck):} every operator
round-trips intermediate data through DDR. \emph{Solution~2:
On-Chip Pipeline Fusion.} Merge adjacent operators into a single
kernel with intermediate data retained on-chip.}
\label{fig:methodology}
\end{figure*}

The 46.2~ms receive latency on the 310B1 is well above the
sub-millisecond budget of 5G~NR; on the GPU side, real-time
sub-millisecond neural receivers have been demonstrated on
datacenter-class inference platforms~\cite{wiesmayr2024realtime},
where the relevant comparison is not absolute latency but
latency under matched power envelopes, since edge AI-RAN nodes
operate at one to two orders of magnitude lower power than
those platforms. Closing the absolute latency gap on edge NPUs
is the immediate technical objective for our line of work. The path forward is to move to
higher-performance NPU platforms rather than to further optimize
operator code on the 310B1 itself, where two factors take effect.
First, datacenter-class members of the same architectural family
deliver substantially higher tensor throughput at identical
programming interface, so the existing operator code benefits
from this throughput increase without rewriting. Second, and more
fundamentally, the set of NPU-native primitives is expected to
broaden on higher-end platforms, with hardware acceleration for
operations such as fused attention, sparse computation, and
fused normalization that emerging AI workloads demand. To the
extent that these richer primitives appear in future NPU
generations, baseband operators that currently rely on manual
composition from Cube and Vector alone may benefit from them as
well, allowing the baseband chain to inherit hardware advances
driven by AI evolution.

A single NPU substrate also supports the other two pillars of
AI-RAN. Traditional and learned operators are expressed through
the same Cube and Vector engines using the same AscendC
interface, so individual operators can be progressively replaced
by neural-network implementations within the existing operator
graph, supporting AI-for-RAN; channel estimation, equalization,
and demapping are the natural first
targets~\cite{cammerer2023nrx,abdollahpour2025}. The inference
capability used here for baseband concurrently supports edge AI
services delivered through the RAN itself, realizing AI-on-RAN.
Future work extends this paradigm to higher protocol layers,
where the scalar-heavy nature of upper-layer processing argues
for hybrid NPU-CPU partitioning rather than a pure-NPU
implementation.

A distinctive research opportunity opened by NPU-based AI-RAN,
enabled by the kernel-level access nCUDA-class interfaces
provide, lies in on-chip orchestration once sufficient AI~Cores
host both workload classes concurrently. NPU on-chip memory is
explicitly programmer-managed through DMA, and AI~Cores are
allocated through software scheduling, allowing predictable
allocation and per-frame reconfiguration. The challenge is the
joint optimization this exposes: a scheduling framework that
allocates cores and tiling parameters across baseband and AI
workloads under shared on-chip memory and baseband latency
constraints, elevating AI-RAN from passive infrastructure
sharing to active compute-communication orchestration within a
single chip.

\section*{Methods}

The mapping presented here is the basic approach for taking
baseband operators onto the matrix and vector engines that
mainstream NPUs share: linear operators are restored to explicit
matrix-multiplication form to land on the matrix engine, while
nonlinear operators are recast as pure batch arithmetic to land
on the vector engine, and adjacent operators are fused on-chip
to amortize data movement. We instantiate this approach on the
Ascend~310B1, the lowest-performance member of the Ascend edge
family, since a successful demonstration on the most constrained
platform constitutes the strongest available evidence that the
approach is viable for the NPU family. The same basic approach
extends to higher-performance NPU platforms, which can
additionally leverage richer primitives and larger on-chip
memory to refine the implementation further; the description in
the remainder of this section reports the implementation as
carried out on the 310B1.

\subsection*{Reshape into matrix and arithmetic}

This subsection addresses the operator-shape mismatch
(Challenge~1 in Fig.~\ref{fig:methodology}) by mapping baseband
operators onto the NPU's matrix and vector engines: linear
operators are restored to explicit matrix-multiplication form to
land on the Cube engine, while nonlinear operators are recast as
pure batch arithmetic to land on the Vector engine.

\subsubsection*{Linear operations: restoring matrix-multiplication form}

\begin{table*}[!t]
\centering
\caption{\textbf{Representative Baseband Operators: CPU/DSP vs.\
NPU-Native Implementations.}}
\label{tab:operator_mapping}
\renewcommand{\arraystretch}{1.3}
\small
\begin{tabular}{@{}llclc@{}}
\toprule
\textbf{Operator} & \textbf{CPU/DSP Form} & \textbf{Engine} & \textbf{NPU-Native Form} & \textbf{Transformation} \\
\midrule
\multicolumn{5}{@{}l}{\textit{Linear operations $\to$ Cube engine}} \\[2pt]
FFT / IFFT           & Butterfly                   & Cube         & Batched DFT matmul                  & Restore matrix    \\
RRC filtering        & Tap-by-tap FIR              & Cube         & Toeplitz matmul                     & Restore matrix    \\
LDPC enc/dec         & Bitwise XOR over GF(2)      & Cube         & Int8 generator matmul + mod\,2      & Restore matrix    \\
LS estimation        & Per-subcarrier division     & Cube         & Pre-computed LS matmul              & Restore matrix    \\
\midrule
\multicolumn{5}{@{}l}{\textit{Nonlinear operations $\to$ Vector engine}} \\[2pt]
Frame sync           & Sequential correlation      & Vector       & Batched energy detect + xcorr       & Vectorized reduce \\
CFO estimation       & Per-symbol CP correlate     & Vector       & Parallel CP autocorrelation         & Vectorized reduce \\
CFO compensation     & Per-sample $e^{j\theta}$    & Vector       & Recursive phasor + angle-sum        & Recast arithmetic \\
QAM modulation       & Constellation LUT           & Vector       & Truncate-scale arithmetic chain     & Recast arithmetic \\
QAM demodulation     & Threshold + bit extract     & Vec + Scl    & Quantize-clamp + scalar bit extract & Recast arithmetic \\
\midrule
\multicolumn{5}{@{}l}{\textit{Adjacent operator pairs $\to$ fused single kernel}} \\[2pt]
LS + ZF equalization & Two independent kernels     & Cube$\to$Vec & Single kernel; on-chip direct pass  & On-chip fusion    \\
\bottomrule
\end{tabular}
\end{table*}

The linear operations that dominate baseband processing
(transforms, filtering, channel coding) are at their mathematical
core matrix multiplications, although decades of optimization for
general-purpose processors have led to alternative
implementations. The Cooley-Tukey butterfly reduces the Discrete
Fourier Transform (DFT) from $O(N^{2})$ to $O(N\log N)$; Finite
Impulse Response (FIR) filtering becomes a tap-by-tap
inner-product loop; Low-Density Parity-Check (LDPC) encoding is
carried out through sparse bitwise-XOR sequences. These
reformulations follow a sound CPU/DSP principle: fewer operations
mean shorter execution. On an NPU the principle no longer holds.
The butterfly's bit-reversed, non-contiguous access pattern, the
scattered tap loop, and the sparse XOR sequence all bypass the
Cube engine because none matches the dense, stride-1
matrix-multiply pattern it is built to accelerate. A carefully
optimized $O(N\log N)$ algorithm may therefore run \emph{slower}
than its $O(N^{2})$ textbook counterpart: the former
underutilizes the hardware, while the latter aligns with its
computational pattern.

The basic idea is to restore each operation to an explicit
matrix form that the Cube engine can execute as a dense matmul;
the specific form differs across operators depending on the
nature of the mismatch. The Cube engine operates on real-valued
matrices, so complex matrix products are decomposed into
real-valued sub-matmuls. For the DFT, we adopt a cos/sin basis
decomposition, factoring the DFT matrix into real cosine and
sine bases and computing the transform as two real matmul calls
combined by an element-wise post-processing kernel; this
replaces the butterfly's bit-reversed access pattern with
stride-1 dense reads, simplifies bias injection (pilot insertion
on transmit, DC-bin handling on receive), and matches the
engine's real-valued data path. RRC pulse shaping, conventionally
implemented as a sliding-window FIR convolution, is rewritten as
a polyphase Toeplitz matmul that materializes the sliding window
as a structured matrix. LDPC encoding, conventionally a sequence
of bitwise-XOR operations over the Galois Field $\mathrm{GF}(2)$,
is lifted into an int8 generator-matrix multiply followed by a
cast and a bitwise-AND mod-2 reduction, trading single-bit logic
for integer arithmetic on the multiply-accumulate array. Even
data subcarrier extraction, conventionally a scalar loop indexing
into the frequency-domain output, is recast as a
permutation-matrix matmul that runs on the Cube engine. Each
case addresses a different mismatch between the conventional
algorithm and the Cube engine: an access-pattern mismatch for
the DFT, a temporal-structure mismatch for RRC, a
computation-type mismatch for LDPC, and a data-movement mismatch
for subcarrier extraction.

Restoring the matrix form is only the first step; the matrix
must also be large enough to fill the Cube engine. A single OFDM
symbol produces a thin matrix that leaves most of the compute
array idle. We therefore concatenate all symbols of a frame
along the row dimension, processing the entire frame as one
large batched matrix multiply. This batching pattern recurs
throughout the chain: whenever a linear operator acts
independently across data blocks, gathering them into a single
large matrix multiply lets the Cube engine treat them as a
single well-shaped workload rather than many under-utilized
small ones. Inner matrix dimensions on AscendC must additionally
be aligned to a multi-byte boundary set by the Cube engine's
data path; payload widths that fall between alignment boundaries
are zero-padded at matrix construction time and trimmed at
output, with negligible overhead relative to the matmul itself.

\subsubsection*{Nonlinear operations: recasting as pure arithmetic}

Nonlinear operations such as modulation, demodulation, and
Carrier Frequency Offset (CFO) compensation appear to depend on
mechanisms the Vector engine lacks: lookup tables, bit-level
shifts, trigonometric function calls. These are standard
implementations on platforms without efficient vector arithmetic.
At their mathematical core, however, these operations are
arithmetic functions of their inputs, most of which can be
expressed in pure arithmetic form (cast, add, multiply, clamp)
that the Vector engine executes natively. The basic idea is to
recast each operation into such a form, with two cases of
differing intervention depth illustrating the practice.

Quadrature Amplitude Modulation (QAM) is conventionally
implemented as a lookup from a packed symbol byte into a
constellation table. On the Vector engine, lookup tables incur
per-element scalar gathers and stall the pipeline. We replaced
the lookup with a short batched chain of cast, scale, and offset
operations executed by the Vector engine, mapping each symbol
byte directly to its in-phase and quadrature levels through
arithmetic alone. QAM demodulation is structurally symmetric: a
single batch quantization recovers the symbol index from the
equalized amplitude. A granularity mismatch remains at the
boundary between LDPC's continuous bit stream and QAM's
per-symbol byte indices; extracting indices on the NPU would
require shifts and masks that leave the Vector engine idle, so
the host CPU instead performs this rearrangement using ARM NEON,
which is empirically faster than the equivalent NPU scalar path
due to the per-row synchronization cost on the NPU scalar unit.
The host boundary is generally where data rearrangement (bit
packing, phase interleaving, stride-N memory patterns) belongs:
vector hardware on the host accommodates these patterns
natively, while the NPU scalar unit pays a synchronization cost
per row.

CFO compensation requires a deeper rewrite of the algorithm
skeleton itself, not just its arithmetic form. A direct
implementation requires an independent complex exponential at
each sample, a per-sample trigonometric evaluation that would
funnel onto the scalar unit. The phase increment between samples
is a known constant, however, so the entire compensation phasor
can be generated on-chip via angle-sum identities. A single
scalar recursion pre-computes a base $\cos/\sin$ vector of one
tile length; each subsequent tile is generated from the base by
a tile-level rotation factor $(\cos\theta_t, \sin\theta_t)$ via
the angle-sum identity, realized as a few Vector multiply-add
operations per tile. The compensation phasor is then multiplied
element-wise with the input signal, also on the Vector engine.
The trigonometric evaluation is thereby transformed from a
per-sample scalar burden into a one-time scalar recursion
followed by batch vector arithmetic, with the entire chain
remaining on the Vector engine across all available cores.

\subsection*{On-chip pipeline fusion}

This subsection addresses the data-movement bottleneck
(Challenge~2 in Fig.~\ref{fig:methodology}) by merging adjacent
operators into a single kernel that retains intermediate data
on-chip throughout. The Cube and Vector engines complete their
work quickly, but moving results to off-chip global memory and
fetching them back for the next operator can take longer than
either computation itself, an effect particularly pronounced in
baseband processing, where each OFDM frame carries large
intermediate tensors between adjacent operators.

The prerequisites for fusion are simple: the operators must
execute sequentially, and the intermediate result must fit in
on-chip local memory after tiling. LS channel estimation
followed by Zero-Forcing (ZF) equalization, a representative
OFDM receiver pair, illustrates the pattern. In the fused kernel
the Cube engine produces the channel estimate via a pilot-by-LS
matmul, and the Vector engine immediately consumes it for
complex-valued ZF division: the estimate is produced, used, and
discarded entirely on-chip within one kernel invocation. The
same fusion principle extends beyond pairs of neighboring
kernels into stride-level fusion across the chain. Wherever
adjacent kernels can agree on a shared output and input layout,
the layout-conversion kernel that would otherwise sit between
them is no longer needed: pulse-shape filtering on receive can
write directly into the FFT input layout, dissolving
cyclic-prefix removal into a stride offset rather than a
separate kernel; modulation can write directly into the
transform input stride, removing the host-side padding step.
Each such stride-level fusion removes a kernel from the chain or
eliminates a host-device round-trip from it, incrementally
tightening the pipeline. Table~\ref{tab:operator_mapping}
summarizes the per-operator transformations across the OFDM
transceiver chain.

\subsection*{Experimental setup and OFDM system parameters}

The mapping requires kernel-level programming access to both
matrix and vector engines through an nCUDA-class interface;
AscendC provides this on Huawei's Ascend NPUs, while other
mainstream NPU platforms currently expose only inference
runtimes or partial engine access.

The verification platform consists of two Orange Pi AI Pro
single-board computers, each equipped with an Ascend~310B1 NPU,
paired with Ettus Universal Software Radio Peripheral (USRP)~X300
software-defined radios and horn antennas, forming a
Single-Input Single-Output (SISO) link at 3.0~GHz. The transmit
chain on Board~A LDPC-encodes random information bits, maps them
to 64-QAM constellation points, assembles them into OFDM symbols
via IFFT with pilot insertion and CP addition, and pulse-shapes
the result through RRC filtering with four-times oversampling. A
Zadoff-Chu synchronization sequence is prepended before
transmission. The receive chain on Board~B captures raw IQ
samples and inverts this process: frame synchronization via
Zadoff-Chu correlation, CFO estimation from CP autocorrelation
followed by time-domain compensation, RRC matched filtering with
downsampling, OFDM demodulation via FFT, LS channel estimation
with ZF equalization, 64-QAM demodulation, and LDPC decoding. The
key OFDM parameters used throughout this work, chosen consistent
with 5G New Radio physical-layer design
principles~\cite{lin2018_5gnr,cacciapuoti2021_5gnr}, are a 256-point FFT
with 16-sample cyclic prefix, 220 data and 16 pilot subcarriers,
64-QAM modulation, rate-1/2 LDPC$(512,256)$ coding, RRC pulse
shaping with roll-off $\beta = 0.35$ and four-times oversampling,
and a Zadoff-Chu synchronization sequence of length $N=256$
($u=1$), all carried at a 5~MHz sampling rate over a 3.0~GHz RF
carrier.

\section*{Data availability}

The experimental measurement data and cross-platform benchmark
results supporting this study are openly available at Zenodo:
\url{https://doi.org/10.5281/zenodo.20107954}. The dataset
includes 200 over-the-air capture frames (complex64 IQ samples),
the transmit-side ground-truth bits, per-frame BER statistics,
end-to-end and per-operator latency measurements across four
platforms, and the complete OFDM system configuration.

\section*{Code availability}

The AscendC operator implementations and host-side processing
code used in this study are publicly available at
\url{https://github.com/refresh3939/AI-RAN-on-NPU}.

\section*{Competing interests}

The authors declare no competing interests.

\section*{Funding Statement}

This paper was funded in part by Jiangsu Major Project on Fundamental Research 
(Grant No.: BK20243059), in part by Gusu Innovation Project 
(Grant No.: ZXL2024360), in part by High-Tech District of Suzhou City 
(Grant No.: RC2025001), in part by Interdisciplinary Disciplines Breakthrough 
Plan of the Ministry of Education of China (No. JYB2025XDXM118), in part by the 
Natural Science Foundation of China (Grant No.: 62301122), and Xiaomi Young 
Scholar Award (Grant No.: Not applicable).

\end{document}